\documentclass[12pt]{article}
\begin{document}

\addtolength{\baselineskip}{0.5\baselineskip}

\title{\textbf{Theoretical Study of Quantum Scattering Processes for Diatomic Hydrogen $(^{2}S)$ and
Oxygen $(^{3}P)$ Complex}}
\author{Liqiang Wei, Michael J. Jamieson and Alexander Dalgarno\\
Institute for Theoretical Atomic, Molecular and Optical Physics\\
Harvard University, Cambridge, MA 02138} \maketitle

\begin{abstract}
\vspace{0.05in}
 We present a quantum mechanical study of the diatomic hydrogen $H(^{2}S)$ and oxygen $O(^{3}P)$
 collision and energy transfer for its four molecular symmetry $(X^{2}\Pi, ^{2}\Sigma^{-}, ^{4}\Pi, ^{4}\Sigma^{-})$,
 which is important for the investigation of many processes of astrophysical and chemical interests including the
 one on molecular cooling, trapping or Bose-Einstein condensation. We compute the rovibrational spectra
 for the $(X^{2}\Pi)$ state and the resulting bound states are in an excellent agreement with experimental
 data for the energy range lower than the dissociation threshold.  We calculate the phase shifts of its partial-waves,
 the total cross section, and the differential cross section. They are all well-structured because of the shape of its
  potential curve. We do the similar studies for the other three dissociative states
   $(^{2}\Sigma^{-}, ^{4}\Pi, ^{4}\Sigma^{-})$. Finally, we have also decided the thermal rate coefficients for the
    hydrogen and  oxygen collision for its four individual states.
\end{abstract}

\vspace{0.35in}
\section{Introduction}
 The atom and atom collisions are the simplest nevertheless among the most important systems for
 quantum molecular scattering investigation. On one hand, they serve as a prototype for the understanding
  of quantum scattering process because they are the only real molecular systems
 at present time that a complete picture of the different aspects
of the quantum scattering can been seen from a computational
simulation, and that a strict test with the experimental
observations can be made ~\cite{mott,brandsen}. On the other hand,
good understanding and accurate calculation of related quantities
for the diatomic scattering are essential to the exploration of
many other important fields. This includes the study of laser
cooling, magnetic trapping and Bose-Einstein condensation of
molecules ~\cite{bohn,ye}. To these ends, we have chosen the
system of oxygen-hydrogen collision for our present study,
 \begin{equation}
  O^{*} + H \rightarrow O + H^{*}.
\end{equation}
 On the practical or application side, this reaction constitutes the basic or primary step for many
  complicated processes for combustion, chemical reactions, and atmospheric or astrophysical
 sources [5-30].
 For example, it is one of the main reactions occurring in the nonthermal escape of
 the hydrogen from Venus and Mars, where $O^{*}$ is produced due to the dissociative recombination of
 $O^{+}_{2}$ in the Martian atmosphere
 ~\cite{alex1,hodges1,shizgal1,shizgal2}. Another example is that the
 collisional hydrogen and oxygen has been investigated to
 illustrate the use of infrared laser pulses to manipulate the
 collision pairs~\cite{schmidt1,schmidt2}.

 In this paper, our object is to present a relatively detailed and systematic study of the collision for the
 system (1). In Section 2, we first show the potential energy curves of the first four states
  of the collision (1) and decide the bound states for the ground state due to the existence of an attractive
  potential well. We then present the fundamental theory for the investigation of their quantum scattering process.
  In Section 3, we demonstrate the various aspects of the scattering processes. We first show the behavior of
  most fundamental quantity - phase shift - for the quantum scattering and study its variation in terms of the
  partial waves or collision energy. We then calculate the total cross sections and the differential cross sections
  and study their relations with the change of energies or scattering angles. We also compute the thermal rate
  coefficients which is the average of the reactive flux over the Maxwell-Boltzmann distribution. The final
  Section contains a discussion and summary.

\vspace{0.35in}
\section{Potential Energy Curves and Scattering Theory}

\subsection{Potential energy curves}

We first present the potential energy curves for the collisional
$O(^{3}P)+H(^{2}S)$ system. It is well-known that the dynamic
features, especially the resonance, is decided by or even sensible
to the details of the shape for the curves~\cite{wei}. In
addition, the previous calculations
~\cite{hodges1,shizgal1,shizgal2} for the excited states
$(^{2}\Sigma^{-},\ ^{4}\Pi,\ and\ ^{4}\Sigma^{-})$ use the
potential data which are approaching the asymptotic region in the
positive direction which is in violation of the physical law of an
attractive dispersion force at a large
distance~\cite{alex2,alex3,alex4,alex1,grigo}. Therefore, we
decide to use the one from the calculation of Yarkony without an
inclusion of any fine-structure corrections such as the spin-orbit
interaction ~\cite{yarkony1,yarkony2}. We do the analytical
fitting and require that the asymptotical potentials satisfying
the negative or inverse $6$th-power of internuclear distance.
Figure $1$ displays the first four states $(X^{2}\Pi,
^{2}\Sigma^{-}, ^{4}\Pi, ^{4}\Sigma^{-})$ of the system (1) with
the energy zero chosen to be at the asymptotes. The curve for the
$X^{2}\Pi$ state is the typical one of a ground state for a
diatomic molecule, featuring an attractive potential well at a
stable configuration. The other three are the usual repulsive
potentials seen for the excited states.

 The molecular spectra of the $OH$ system is still of fundamental
 interest~\cite{mantz,herzberg}. In addition to the continuing investigation of its whole range of
 the spectra, the $OH$ maser is considered to be
 a means for detecting the interaction between the supernova
 remnant and molecular cloud which is important for understanding
 the formation processes of new stars~\cite{alex5,melen,wardle}.
 For the ground state $(X^{2}\Pi)$ potential energy curve, we have determined the
 bound states associated with the attractive potential energy well for the energy
 below the dissociation threshold.  We do the calculation by utilization of the
 $LEVEL\ 7.5$ program~\cite{LeRoy}. The Hamiltonian includes the standard contributions for a
 diatomic molecule. The resulting rovibrational energy levels constitute
 a banded structure with each vibrational quantum number $n$ associated with a bunch of rotational
 sublevels $\{j\}$. The table $1$ shows the computed results for vibrational quantum number
 up to $16$ and when the $j$ equal to $0$ or $1$. When compared to
 the experimental observation~\cite{mantz,herzberg}, they have an excellent
 agreement except for the states near the dissociation threshold
 where we expect a big derivation from the behaviors for the ideal models.

\subsection{Scattering Theory}

We employ the usual approach of  partial wave expansion to study
the atom-atom collisions where both the incident and the scattered
waves are analyzed in terms of the components of spherical
waves~\cite{mott,brandsen}. Since only the phase shifts
$\delta_{l}$ of the outgoing partial waves are changed, they are
the central quantity that need to be calculated. They contain all
the dynamic information for the scattering process. The phase
shifts are determined by the requirement that the inside and
outside scattering waves and their derivatives are continuous at
the boundary point $R$ where the interaction potential $V(r)$
between two atoms vanishes. They are given by the formula
\begin{equation}
 \tan \delta_{l} = \frac{u(R)k\tilde{j_{l}}^{'}(kR)-u^{'}(R)\tilde{j_{l}}(kR)}
 {u(R)k\tilde{n_{l}}^{'}(kR) - u^{'}(R) \tilde{n_{l}}(kR)},
\end{equation}
where
\[   k = \sqrt{\frac{2mE}{\hbar^{2}}} \]
and $\tilde{j_{l}}$ or $\tilde{n_{l}}$ are the Ricatti-Bessel or
Ricatti-Neumann functions. The function $u(R)$ is the radial
scattering wave obtained by numerically solving the radial
Schr$\ddot{o}$dinger equation from $R=0$ for the given central
potential. Once the phase shifts are known, the differential or
the total cross sections, for instance, can be easily calculated
as follows
\begin{equation}
     \frac{d\sigma_{tot}}{d\Omega} = |f(\theta)|^{2},
\end{equation}
where
\begin{equation}
    f(\theta) = \frac{1}{k}\sum_{l=0}^{\infty}(2l+1)e^{i\delta_{l}}\sin\delta_{l}
    P_{l}(\cos\theta)
\end{equation}
and
\begin{equation}
 \sigma_{tot} = \frac{4\pi}{k^{2}}\sum_{l=0}^{\infty}(2l+1)\sin^{2}\delta_{l}.
\end{equation}

\vspace{0.35in}
\section{Calculation Results}

\subsection{Partial wave analysis}
 We first do the partial wave analysis of the ground state $(X^{2}\Pi)$ for the $O+H$ collisional system.
  Figure $2$ illustrates the variation of the calculated phase shifts for the $s$ and $p$ waves in terms of the
  total collision energy. We observe the expected phase change when the angular momentum quantum number
  changes by one unit. For the energies computed, there are many partial phases contributing to the total
  cross section for the scattering process. In addition, they have the trend for approaching the zero when
  seen from the lower energy side.

 For the excited states, the partial wave analysis can be carried out in the same way.
  Figure $3$ is related to the change of the $s$ and $p$ waves in terms of the collision energy for the
  $^{4}\Sigma^{-}$ state. We notice the similar behaviors to those for the $X^{2}\Pi$ state.

\subsection{Total and differential cross sections}

Figure 4 depicts the total cross sections for the four electronic
states of the hydrogen $(^{2}S)$ and oxygen $(^{3}P)$ collisional
system with different symmetries. The energy range is from $0.27\
eV$ to $15\ eV$.
As expected, they are a decreasing function of the total energy
except at the points where resonance occurs for the ground state.
It is due to the existence of a potential well for this state
$(X^{2}\Pi)$. In addition, the magnitude of the total cross
section for the $X^{2}\Pi$ state is significant larger than those
for the other three repulsive states. The latter states show the
similar behaviors.

 We also compute the differential cross sections for the collision (1) with different orientations,
  which is a key quantity, for instance, for the study of escaping issues of hydrogen and deuterium in the planetary
 exosphere. Figures 5 displays the calculated results for the $X^{2}\Pi$ state for the energy
 equal to $0.2\ eV$ or $3.0\ eV$, respectively. We see that for the low energy there is a very large probability
 for forward scattering $(\theta=0)$ and also a noticeable amount for the head-on collision $(\theta=\pi)$.
 When energy increases, however, the chance for the forward scattering persists but the one for the head-on
 collision diminishes. The other states have the similar situations but they have no
  probability for head-on collision for any values of energies. This is shown in Figure 6.

\subsection{Reaction rate coefficients}
  The thermal reaction rate coefficients at temperature $T$ are defined as
  follows ~\cite{alex6,alex7,alex8}
\begin{equation}
 k(T) = <v\sigma> = \left(\frac{8}{\pi\mu\beta}\right)^{1/2}
 \int^{\infty}_{0} \left(\beta E\right) e^{-\beta E}\sigma(E) d\left(\beta E\right),
\end{equation}
 which are the Maxwell-Boltzman average over the reactive flux. The
reactive flux is the relative velocity $v$ for the collision times
the total cross section $\sigma$ calculated before. The $E$ is the
relative translation energy for the diatomic system given by $ E =
\frac{1}{2} \mu v^{2} $, and the $\beta$ is the inverse of the
Boltzman constant $k_{B}$ times the absolute temperature $T$. The
calculated temperature dependence of the reaction rate coefficient
for the $^{4}\Pi$ state is demonstrated in Figure $7$. The
temperature range chosen is of planetary interest. We see that the
calculated result shows an increasing function of the temperature.
The other states have the similar behavior.

\vspace{0.35in}
\section{Discussions and Summary}

 We have decided the bound states associated with the existence of an attractive
 potential well for the ground state $(X^{2}\Pi)$. The computed rovibrational
 energy levels are in an excellent accord with the experimental data for the lower energy range.
 However, when the energy increases up to the direction of the dissociation threshold, the
 difference is accumulating and becoming very large. This indicates that the standard models
 for treating the rovibrational motion of diatomic molecules
 are no longer valid for the higher energy range.

  The phase shifts are the most fundamental quantity which we have
  computed for the scattering system (1). They shows the expected
  behaviors in terms of the variation of the angular momentum or
  the total collision energy. Our calculated total cross sections for the four individual states
  are similar to the ones obtained by one of the authors~\cite{alex1}, and also in the similar range
  of the magnitudes to some other calculations as shown in the references~\cite{hodges1} and~\cite{shizgal2}.
   However, there are still some significant differences when compared to all these previous study. For example,
  the new total cross section for the ground states is significantly separated from all other
  three dissociation states.  In the meantime, the dissociation
  states all have the similar magnitudes in the total cross
  section. These should be more reasonable results.

 The calculated differential cross sections as demonstrated in Figures 5 and 6 yield some useful information.
 For instance, for the ground state and with a smaller total energy, the forward scattering is a
 preferential one with some probability of backward scattering.
 When energy increases, however, the collision becomes exclusively forward
 and there is no chance for the head-on collision. This is consistent with
 the picture of the potential energy curve for the ground state.
 The existence of the potential energy well results in the quantum
 mechanical interference effect and therefore the splitting of the particle wave. With
 an increasing total energy or larger impact parameter, a classical description becomes
 valid for the the collision. Nevertheless, in contrast to the one for the ground state, the other three
dissociative states have a dominate forward scattering. The
diatomic system collides classically and transfers the energy to
each other. This is also in accord with the shape of the potential
curves as well as an isotropic diatomic interaction.

 Finally, we have determined the rate coefficients for the four
 individual states of the collisional hydrogen $(^{2}S)$ and oxygen
 $(^{3}P)$ system. They are all the increasing function of the
 temperature.

  In conclusion, the collisional hydrogen $(^{2}S)$ and oxygen $(^{3}P)$ complex is one of the
 simplest diatomic systems with many important and practical applications. From the
 investigation presented in this paper, we have gained a better understanding
 of its scattering processes. Furthermore, our computation has demonstrated how significant the shape of
  the potential energy curves is in determining the dynamic features of a molecular system. The range
 of energies being studied is also an important consideration.
  Therefore, a dynamic theory or approach, incorporating quantum and classical mechanics
  transition, will be a powerful tool in the study of molecular dynamics for large systems.

\vspace{0.45in} \noindent

{\bf \Large{Acknowledgment}}

\vspace{0.15in} \noindent

This work was supported by the National Science Foundation through
a grant to Professor Dalgarno, and a grant to the Institute for
Theoretical Atomic, Molecular and Optical Physics at Harvard
University and Smithsonian Astrophysical Observatory.

\newpage

  \begin{center}
{\bf Table Caption}
\end{center}

\mbox{}\\
\noindent {\bf Table 1.} The energies (in unit of $eV$) of the
rovibrational bound states of diatomic molecule $OH$ with angular
momentum $j=0$ or $1$ associated with the attractive potential
well for its ground state $X^{2}\Pi$ .

\begin{table}[p]
\begin{center}
  \begin{tabular}{|l|l|l|l|}  \hline\hline
 $n,j$ & \ \ \ \ \ $E(n,j)$ & $n,j$ & \ \ \ \ \ $E(n,j)$ \\  \hline\hline
  0, 0  &  -4.349280532476708 & 0, 1 & -4.344677732905585 \\ \hline
 1, 0 & -3.9007571837193575 & 1, 1 & -3.8963314535284636  \\ \hline
 2, 0 & -3.4743285871919105 & 2, 1 & -3.470079566811745 \\ \hline
 3, 0 & -3.069862942074269  & 3, 1 & -3.065790705898521  \\ \hline
 4, 0 & -2.68728756573168  & 4, 1 &  -2.683392721308816  \\ \hline
 5, 0 & -2.326599978374491 & 5, 1 &  -2.3228837532003874 \\ \hline
 6, 0 & -1.9878821494497058 & 6, 1 & -1.9843465893508291  \\ \hline
 7, 0 & -1.6713195300265837 & 7, 1 & -1.6679677223410522  \\ \hline
 8, 0 &  -1.3772272393753788 & 8, 1 & -1.3740636477245691 \\ \hline
 9, 0 & -1.1060875469879576  & 9, 1 & -1.1031184700375904 \\ \hline
 10, 0 & -0.8586060398134588 & 10, 1 & -0.8558403676094428 \\ \hline
 11, 0 & -0.6358007830952933 & 11, 1 & -0.6332512865543456 \\ \hline
 12, 0 & -0.43915758000136823 & 12, 1 & -0.4368432791101333 \\ \hline
 13, 0 & -0.27104694825436975 & 13, 1 & -0.2690016425458286 \\ \hline
 14, 0 &  -0.1378520028115584 & 14, 1 & -0.1361891922581645 \\ \hline
 15, 0 & -0.05391304486808863 & 15, 1 & -0.05271912534137973 \\ \hline
 16, 0 & -7.919976994138954E-3 & 16, 1 & -7.246986879999613E-3  \\  \hline\hline
   \end{tabular}
\end{center}
\end{table}

\newpage

  \begin{center}
{\bf Figure Captions}
\end{center}

\mbox{}\\
\noindent {\bf Figure 1.} The fitted potential energy curves for
the four states $(X^{2}\Pi,^{2}\Sigma^{-},^{4}\Pi,^{4}\Sigma^{-})$
of the diatomic molecule $OH$.

\mbox{}\\
\noindent {\bf Figure 2.} The phase shifts of the $s$ and $p$
waves of the ground state $(X^{2}\Pi)$ for the diatomic molecule
$OH$ and their variation in terms of the collision energy.

\mbox{}\\
\noindent {\bf Figure 3.} The phase shifts of the $s$ and $p$
waves of the excited state $^{4}\Sigma^{-}$ for the diatomic
molecule $OH$ and their variation in terms of the collision
energy.

\mbox{}\\
\noindent {\bf Figure 4.} The total cross sections (in unit of
$\dot{A}^{2}$) for the four states
 $(X^{2}\Pi,^{2}\Sigma^{-},^{4}\Pi,^{4}\Sigma^{-})$ of the diatomic
molecule $OH$ in the energy range from $0.27\ eV$ to $15\ eV$.

\mbox{}\\
\noindent {\bf Figure 5.} The differential cross sections (in
atomic units) for the ground state $(X^{2}\Pi)$ of the diatomic
molecule $OH$ at the energies $0.2\ eV$ or $3.0\ eV$.

\mbox{}\\
\noindent {\bf Figure 6.} The differential cross sections (in
atomic units) for the four states
$(X^{2}\Pi,^{2}\Sigma^{-},^{4}\Pi,^{4}\Sigma)^{-}$ of the diatomic
molecule $OH$ at the energy $0.2\ eV$.

\mbox{}\\
\noindent {\bf Figure 7.} The temperature-dependence of the
reaction rate coefficient (in unit of $cm^{3}s^{-1}$) for the
state $^{4}\Pi$ of the diatomic molecule $OH$.

\end{document}